\documentclass[aip,pop,reprint]{revtex4-1}
\usepackage[latin9]{inputenc}
\usepackage{textcomp}
\usepackage{amsmath}
\usepackage{amssymb}
\usepackage{graphicx}
\usepackage{esint}

\DeclareRobustCommand{\greektext}{%
  \fontencoding{LGR}\selectfont\def\encodingdefault{LGR}}
\DeclareRobustCommand{\textgreek}[1]{\leavevmode{\greektext #1}}
\DeclareFontEncoding{LGR}{}{}
\DeclareTextSymbol{\~}{LGR}{126}
\newcommand{\lyxmathsym}[1]{\ifmmode\begingroup\def\b@ld{bold}
  \text{\ifx\math@version\b@ld\bfseries\fi#1}\endgroup\else#1\fi}

\begin{document}

\title{Production and dynamics of positrons in ultrahigh intensity laser-foil
interactions }

\author{I.~Yu.~Kostyukov}

\email{kost@appl.sci-nnov.ru}
\author{E.~N.~Nerush}

\affiliation{Institute of Applied Physics, Russian Academy of Science, 46 Uljanov
str., 603950 Nizhny Novgorod, Russia}
\begin{abstract}
The electron-positron pair production accompanying interaction of
a circularly polarized laser pulse with a foil is studied for laser
intensities higher than $10^{24}$W cm$^{-2}$. The laser energy penetrates
into the foil due to the effect of the relativistic hole-boring. It
is demonstrated that the electron-positron plasma is produced as a
result of quantum-electrodynamical cascading in the field of the incident
and reflected laser light in front of the foil. The incident and reflected
laser light makes up the circularly polarized standing wave in the
reference frame of the hole-boring front and the pair density peaks
near the nodes and antinodes of the wave. A model based on the particle
dynamics with radiation reaction effect near the magnetic nodes is
developed. The model predictions are verified by 3D PIC-MC simulations. 
\end{abstract}


\maketitle

\section{Introduction}

Ultrahigh intensity laser-matter interaction attracts much attention,
first of all, due to the fast development in laser technology \cite{Michigan2008,Mourou2006}.
At extremely high laser intensity the quantum-electrodynamical (QED)
effects start to play a key role. Among them are: photon emission
by electrons and positrons with strong recoil, photon decay in strong
electromagnetic field with electron-positron pair creation (Breit-Wheeler
process), Bethe-Heitler process, trident process, etc \cite{Marklund2006,Piazza2012}.
The laser-matter interaction in the QED-dominated regime leads to
manifestation of new phenomena like prolific production of gamma-rays
and electron-positron pairs \cite{Nerush2007,Ridgers2012,Kirk2013,Bashinov2014,Brady2014,Nerush2014,Nerush2015},
laser-assisted QED cascading \cite{Bell2008,Fedotov2010,Nerush2011,Bulanov2013,Bashmakov2014,Gelfer2015,Vranich2015,Jirka2016},
radiation trapping of the charged particles \cite{Lehmann2012,Ji2014prl,Gonoskov2014,fedotov2014,Kirk2016}
etc.

In this paper we focus on laser-plasma interaction in the hole-boring
(HB) regime when the light pressure pushes plasma inside the target
\cite{Kruer1975,Wilks1992}. The hole-boring front can be introduced
as a plasma-vacuum interface propagating towards the target. The front
separates the vacuum region from the high density plasma. The HB front
structure is as follows. The laser pressure pushes the electrons ahead
thereby forming the sheath with the unshielded ions and the thin,
dense electron layer. Reflection and absorption of the laser light
by the electron layer provides efficient laser pressure. The charge
separation generates strong longitudinal electric field that, on the
one hand, accelerates the ions towards the target and, on the other
hand, suppress the electron acceleration by the laser pressure. Laser
radiation and the plasma ions mostly contribute into the energy-monetum
budget. The HB front velocity can be derived from the equation for
the energy-monetum flux balance \cite{Kruer1975,Schlegel2009,Robinson2009}

\begin{eqnarray}
v_{HB} & = & \frac{c}{1+\mu},\label{vhb}
\end{eqnarray}
where

\begin{eqnarray}
\mu & = & \frac{1}{a_{0}}\sqrt{\frac{Mn_{i}}{mn_{cr}}},\label{mu}
\end{eqnarray}
$a_{0}=eE/(mc\omega_{L})$ is the normalized laser field strength,
$n_{cr}=m\omega_{L}^{2}/\left(4\pi e^{2}\right)$ is the critical
plasma density, $n_{i}$ is the density of the plasma ions, $\omega_{L}$
is the laser frequency, $M$ is the ion mass, $c$ is the speed of
light, $m$ and $e>0$ are the electron charge and mass, respectively.
It follows from Eqs.~(\ref{vhb}) and (\ref{mu}) that the HB front
velocity increases with increasing of the laser intensity and decreasing
of the plasma density. The electrons in the laser field can emit high
energy photons and if the laser intensity is high enough then the
portion of the laser energy converted into the gamma ray energy is
large \cite{Nerush2014} so that the fluxes of the emitted gamma-photons
has to be taken into account in the energy-monetum flux budget \cite{Nerush2015}.
It is demonstrated \cite{Nerush2015,Capdessus2015} that the efficient
generation of gamma-rays reduces the laser reflection and the HB front
velocity.

Another effect accompanying the ultrahigh intensity laser-solid interaction
is electron-positron pair creation \cite{Ridgers2012,Kirk2013,Nerush2015}.
The pairs can be created because of Breit-Wheeler process. Avalanche-like
production of electron-positron pairs and gamma photons is possible
at QED cascading \cite{Bell2008,Fedotov2010}. A cascade develops
as a sequence of elementary QED processes: photon emission by the
electrons and positrons in the laser field alternates with pair production
because of photon decay. A cloud or \textquotedblleft{}cushion\textquotedblright{}
of pair plasma in the laser pulse in front of the target has been
observed in numerical simulations \cite{Ridgers2012}. As the pair
number becomes great, there is back reaction of the self-generated
pair plasma on the laser-solid interaction. It has been demonstrated
\cite{Ridgers2012,Nerush2015} that the produced pair plasma dramatically
enhances laser field absorption and gamma-ray emission thereby reducing
the HB front velocity. The pair motion in the combined laser and plasma
fields with radiation reaction is rather complex that makes analytical
treatment of pair plasma kinetics difficult. The analytical model
for pair cushion in the nonlinear regime when the reflection of the
laser pulse is strongly suppressed by the self-generated pair plasma
has been recently proposed \cite{Kirk2013}. 

In our work we study the pair production in the regime when the number
of the produced pairs is not sufficient to suppress laser reflection
and to affect the laser-foil interaction. This is the case, for example,
for interaction between extremely intense laser pulse and thin foils
or for early stage of the laser interaction with thick solid target.
The results of three dimensional particle-in-cell Monte Carlo (3D
PIC-MC) simulations demonstrating the HB effect at interaction between
a circularly polarized laser pulse and a foil are shown in Fig.~1.
PIC-MC simulations including emission of hard photons and electron-positron
pair production allow us to analyze the HB process at extremely high
intensities. A similar numerical approach has been used in a number
of works (see, e. g. \cite{Ridgers2012,Vranich2015}). To distill
the physics of pair production and pair dynamics we consider extremely
intense laser pulses. The simulation box is $17.5\lambda\times25\lambda\times25\lambda$
corresponding to the grid size $670\times125\times125$; the time
step is $0.005\lambda/c$, where $\lambda$ is the laser wavelength.
In the simulation a quasi-rectangular ($11.4\lambda\times23\lambda\times23\lambda$)
circularly polarized laser pulse of intensity $I_{L}=2.75\times10^{24}$
$\mathrm{W/cm^{2}}$ ($a_{0}=1000$, $\lyxmathsym{\textgreek{l}}=1\mathrm{\lyxmathsym{\textgreek{m}}m}$)
and $I_{L}=9.3\times10^{24}$ $\mathrm{W/cm^{2}}$ ($a_{0}=1840$,
$\lyxmathsym{\textgreek{l}}=1\mathrm{\lyxmathsym{\textgreek{m}}m}$)
interacts with a diamond foil ($n_{e}=6n_{i}=1.1\times10^{24}\mathrm{cm^{\lyxmathsym{\textminus}3}}$,
$n_{i}/n_{cr}=158$). The shape of the laser pulse is approximated
as follows 

\begin{eqnarray}
E(x) & \propto & \frac{d}{dx}\left\{ \sin x\cos^{2}\left[\frac{\pi\left(x-x_{s}\right)^{4}}{2x_{s}^{4}}\right]\right\} ,\label{pulse}
\end{eqnarray}
where $x_{s}=5.7\lambda$ (the pulse duration is about $38$ fs).
The pulse has almost constant amplitude in the central area and promptly
decreases at the distance $x_{s}$ from the pulse center. 

\begin{figure}[h]
\centering \includegraphics[width=8cm]{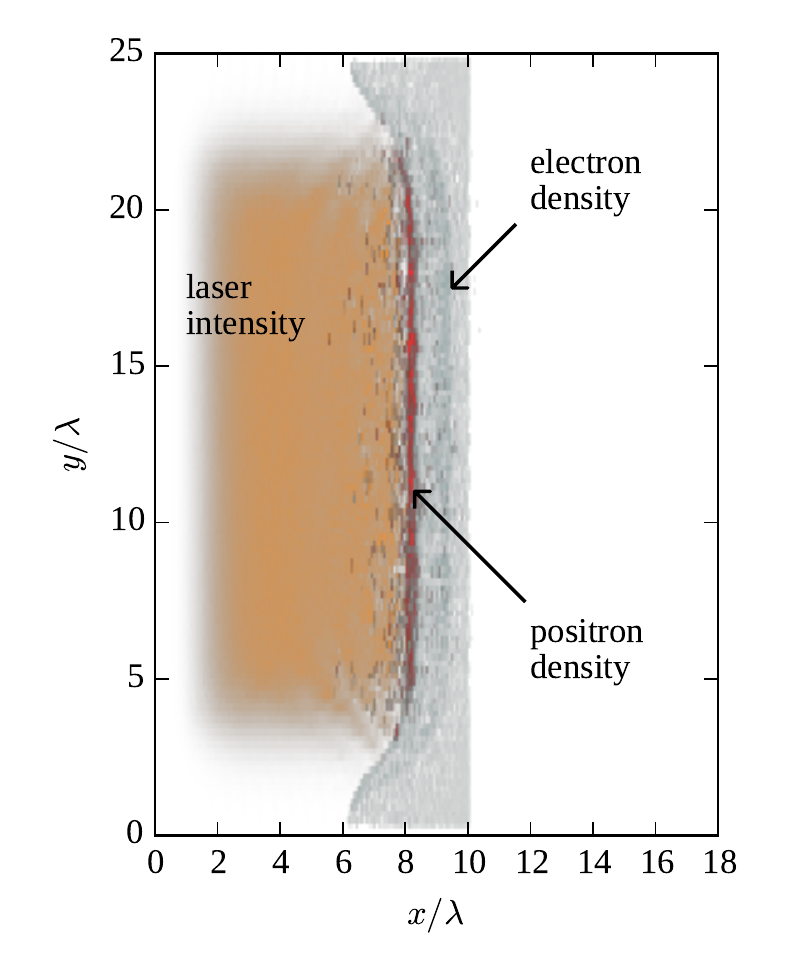} \caption{The distribution of the laser intensity (orange color), the electron
density (gray color) and the positron density (red color) in the $x-y$
plane at $z=0$, $t=4\lambda/c$ for $a_{0}=1840$.}

\label{interaction} 
\end{figure}

\begin{figure}[h]
\centering \includegraphics[width=8cm]{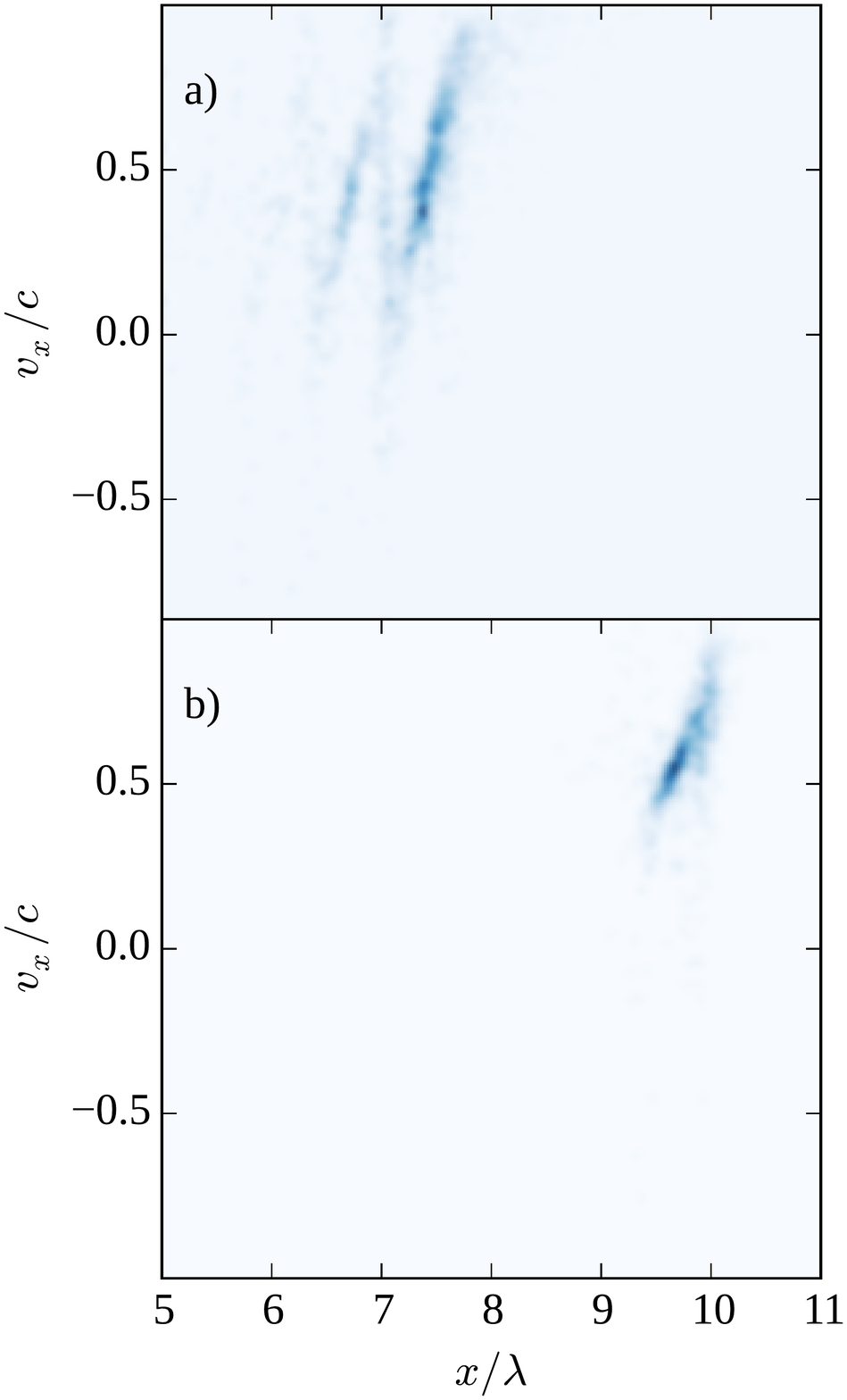} \caption{The positron distribution in the plane $x-v_{x}$, for (a) $a_{0}=1000$,
$t=6.0\lambda/c$ and (b) $a_{0}=1840$, $t=3.0\lambda/c$ .}

\label{xvx-a1000-t6} 
\end{figure}

For $a_{0}=1840$ parameters $\mu=1$ and the velocity of the HB front
is a half of the speed of light. It is seen from Fig.~\ref{interaction}
that the plasma is shifted towards the foil and the thin layer of
electron-positron plasma is produced. The longitudinal phase space
of the positrons produced at the laser-foil interaction is shown in
Fig.~\ref{xvx-a1000-t6} for two values of $a_{0}$ ($a_{0}=1000\mbox{ and }a_{0}=1840$).
In the high intensity regime ($a_{0}=1840$) the positron distribution
is strongly localized in the longitudinal phase space. In the low
intensity regime ($a_{0}=1000$) the positron distribution is sawtooth-like. 

The electron-positron pair production can be roughly divided into
three stages. At the initial stage the electron-positron pairs are
produced from the photons emitted by the foil electrons. This stage
can be described as follows. The laser pulse propagating in the positive
direction of $x$-axis is reflected by the dense electron layer at
the HB front. The layer electrons in a laser field emit a number of
hard photons propagating in the same direction. The HB front outrun
the photons emitted at the large angle to the $x$-axis so after that
the photons move in vacuum region in the field of the incident and
reflected laser radiation. In the HB front reference frame (``HB-frame'')
these photons after escaping from plasma moves in the vacuum region
in the negative direction of $x$-axis in the field of the circularly
polarized standing wave. The structures, which are close to a counter-propagating
wave or a standing wave, are efficient for pair creation \cite{Bell2008,Fedotov2010,Nerush2011}
so that the photons decay and produce electron-positron pairs. In
the second stage the number of pairs becomes so great that the number
of high-energy photons emitted by the pairs exceeds the number of
the external photons emitted by the foil electrons. In this case self-sustained
QED cascade develops in the standing wave. In the final, third stage
there is the back reaction of the self-generated electron-positron
plasma on the laser-foil interaction. The electron-positron plasma
becomes so dense that the significant part of the laser energy is
absorbed by the pairs. In this paper we focus on the first two stages.

\section{Field structure and pair production in the vacuum region}

The incident laser field in the vacuum region can be approximated
by the circularly polarized plane wave propagating along $x$-axis
\begin{align}
\mathbf{E}_{i} & =a_{0}\frac{mc\omega_{L}}{e}\left(0,\cos\varPhi,\sin\varPhi\right),\label{CPinLF}\\
\mathbf{B}_{i} & =a_{0}\frac{mc\omega_{L}}{e}\left(0,-\sin\varPhi,\cos\varPhi\right),\label{CPinLF2}
\end{align}
where $\varPhi=\omega_{L}x/c-\omega_{L}t$ The incident electromagnetic
field in the HB-frame can be calculated with Lorentz transformation
\begin{align}
\mathbf{E}_{i}^{\prime} & =a_{0}\left(0,\cos\left(x^{\prime}-t^{\prime}\right),\sin\left(x^{\prime}-t^{\prime}\right)\right),\label{CPinHBF}\\
\mathbf{B}_{i}^{\prime} & =a_{0}\left(0,-\sin\left(x^{\prime}-t^{\prime}\right),\cos\left(x^{\prime}-t^{\prime}\right)\right),
\end{align}
where prime symbol marks the quantities in the HB-frame. In this Section
we use the dimensionless units, normalizing the time to $1/\omega^{\prime}$
, the length to $c/\omega^{\prime}$ , the momentum to $mc$, and
the field amplitude to $mc\omega^{\prime}/e$, where $\omega^{\prime}=\omega_{L}\gamma_{HB}\left(1-v_{HB}\right)$
is the frequency of the incident wave in the HB-frame and $\gamma_{HB}^{-2}=1-v_{HB}^{2}$.
At the HB front position $x^{\prime}=0$ the boundary condition is
$E_{y}^{\prime}\left(x^{\prime}=0\right)=E_{z}^{\prime}\left(x^{\prime}=0\right)=0$,
where the perfect reflection in the HB-frame is assumed. The reflected
laser radiation can be approximated as follows
\begin{align}
\mathbf{E}_{r}^{\prime} & =a_{0}\left(0,-\cos\left(x^{\prime}+t^{\prime}\right),\sin\left(x^{\prime}+t^{\prime}\right)\right),\label{EriHBF}\\
\mathbf{B}_{r}^{\prime} & =a_{0}\left(0,\sin\left(x^{\prime}+t^{\prime}\right),\cos\left(x^{\prime}+t^{\prime}\right)\right).\label{BriHBF}
\end{align}
In the laboratory frame the reflected wave takes a form
\begin{align}
\mathbf{E}_{r} & =a_{0}\omega_{r}\left(0,-\cos\left(\omega_{r}x+\omega_{r}t\right),\sin\left(\omega_{r}x+\omega_{r}t\right)\right),\label{ErnLF}\\
\mathbf{B}_{r} & =a_{0}\omega_{r}\left(0,\sin\left(\omega_{r}x+\omega_{r}t\right),\cos\left(\omega_{r}x+\omega_{r}t\right)\right),\label{BRinLF}
\end{align}
where

\begin{eqnarray}
\omega_{r} & =\omega_{L} & \frac{1-v_{HB}}{1+v_{HB}},\label{wr}
\end{eqnarray}
 is the frequency of the reflected wave in the laboratory frame. 

As the reflection coefficient is taken to be equal to $1$ in the
HB-frame, the standing wave is generated in the vacuum region:
\begin{align}
\mathbf{E}^{\prime} & =2a_{0}\left(0,\text{\ensuremath{\sin}}x^{\prime}\sin t^{\prime}\text{,}\sin x^{\prime}\cos t^{\prime}\right),\label{Ehbf}\\
\mathbf{B}^{\prime} & =2a_{0}\left(0,\cos x^{\prime}\sin t^{\prime},\cos x^{\prime}\cos t^{\prime}\right).\label{Bhbf}
\end{align}
The wavelength of the standing wave in the HB-frame is
\begin{align}
\lambda^{\prime} & =2\pi=\frac{\lambda}{\gamma_{hb}\left(1-v_{hb}\right)}.\label{lhb}
\end{align}

The probability rate for photon emission by ultra-relativistic electrons
and positrons in an electromagnetic field and the probability rate
for electron-positron pair production via photon decay are given,
respectively, by the formulas \cite{Baier1998}
\begin{eqnarray}
W_{rad} & = & \frac{\alpha a_{S}}{\varepsilon_{e}}\int_{0}^{\infty}dx\frac{5x^{2}+7x+5}{3^{3/2}\pi(1+x)^{3}}K_{\frac{2}{3}}\left(\frac{2x}{3\chi_{e}}\right),\label{Wr1}\\
W_{rad} & \approx & \frac{5\alpha a_{S}}{2\sqrt{3}\pi\varepsilon_{e}}\chi_{e},\;\chi_{e}\ll1,\label{Wr2}\\
W_{pair} & = & \frac{\alpha a_{S}3^{-3/2}}{\pi\varepsilon_{ph}}\int_{0}^{1}dx\frac{9-x^{2}}{1-x^{2}}K_{\frac{2}{3}}\left[\frac{8\chi_{ph}^{-1}}{3\left(1-x^{2}\right)}\right],\label{Wp1}\\
W_{pair} & \approx & \frac{3^{3/2}\alpha}{2^{9/2}}\frac{a_{S}\chi_{ph}}{\varepsilon_{ph}}\exp\left(-\frac{8}{3\chi_{ph}}\right),\;\chi_{ph}\ll1,\label{Wp2}
\end{eqnarray}
where 

\begin{eqnarray}
\chi_{e,ph} & = & \frac{1}{a_{S}}\sqrt{\left(\varepsilon_{e,ph}\mathbf{E}+\mathbf{p}_{e,ph}\times\mathbf{B}\right)^{2}-\left(\mathbf{p}_{e,ph}\cdot\mathbf{E}\right)^{2}},\label{chi_general}
\end{eqnarray}
is the key QED parameter determining the phonon emission ($\chi_{e}$)
and the pair production ($\chi_{ph}$) \cite{Landau4}, $a_{S}=eE_{S}/(mc\omega^{\prime})=mc^{2}/\hbar\omega^{\prime}$
is the normalized QED critical field, $E_{S}=m^{2}c^{3}/(\hbar e)$,
$\varepsilon_{e,ph}$ is the energy of the electron (positron) and
photon, respectively, $\mathbf{p}_{e,ph}$ is the momentum of the
electron (positron) and photon, respectively, $\hbar$ is the Plank
constant. 

The dependence of the pair production probability on $\chi$ is sharp
in the limit $\chi\ll1$. Therefore we can suppose that the most of
the pairs are produced near the points in the spacetime where $\chi$
peaks. If the photon momentum is $\mathbf{p}^{\prime}=\varepsilon_{ph}\left(\cos\alpha,\sin\alpha\cos\beta,\sin\alpha\sin\beta\right)$
then $\chi$ for the circularly polarized standing wave given by Eqs.~(\ref{Ehbf}),
(\ref{Bhbf}) takes a forms in the HB-frame after some trigonometrical
transformations 

\begin{eqnarray}
\chi_{ph} & = & \frac{\varepsilon_{ph}a_{0}}{a_{S}}\sqrt{1-\sin^{2}\phi\sin^{2}\alpha},\label{chi}
\end{eqnarray}
where $\phi=\beta-t^{\prime}$. It follows from Eq.~(\ref{chi})
that $\chi$ peaks at $\phi=\pm\pi n$ or $\alpha=\pm\pi l$, $n,l=0,1,2,...$
and it does not depend on $x^{\prime}$. For given value of $\beta$
there is always the value of $t^{\prime}$ at which $\phi=\pm\pi n$.
If the photon emission is axially symmetrical ($\beta$ is uniformly
distributed from $0$ to $2\pi$) then we can suppose that the pair
number decreases with increasing the distance from the HB front towards
the vacuum region as a result of photon flux attenuation.

\section{Particle motion in a standing circularly polarized wave}

In this Section we study the motion of the electrons and positrons
in the vacuum field in the HB-frame. The filed can be approximated
by the circularly polarized standing wave defined by Eqs.~(\ref{Ehbf})
and (\ref{Bhbf}). Our treatment is based on the classical approach
in order to obtain analytical solutions. In the QED approach the particle
momentum suddenly changes because of recoil effect caused by photon
emission. However even in the limit $\chi\gg1$ the particle energy
much greater than the mean change in its energy because emission of
one photon: $\left\langle \varepsilon_{ph}\right\rangle <I_{rad}(\chi\rightarrow\infty)/W_{rad}(\chi\rightarrow\infty)\approx0.25\ll\varepsilon_{e}$,
where $I_{rad}$ is the total intensity of the photon emission. It
is demonstrated by numerical simulations \cite{Gonoskov2014,Jirka2016}
that the spatial distributions of the electrons calculated in the
classical approach and in QED approach are similar even for extremely
strong electromagnetic fields. 

In the classical approach the positron motion is governed by equations
\begin{align}
\frac{d\mathrm{\mathbf{p}}}{dt} & =\mathbf{F}_{L}-\mathbf{v}F_{R},\label{em1}\\
\frac{d\mathbf{r}}{dt} & =\frac{\mathbf{p}}{\gamma},\label{em2}\\
\mathbf{F}_{L} & =\mathbf{E}+\mathbf{v}\times\mathbf{\mathbf{B}},\\
F_{R} & =\mu a_{S}^{2}\chi_{e}^{2}G\left(\chi_{e}\right),\nonumber \\
\chi_{e}^{2} & =a_{S}^{-2}\gamma^{2}\left[\left(\mathbf{E}+\mathbf{v}\times\mathbf{\mathbf{B}}\right)^{2}-\left(\mathbf{v}\cdot\mathbf{\mathbf{E}}\right)^{2}\right],\label{em3}
\end{align}
where $F_{R}/G\left(\chi_{e}\right)$ is the leading term of the radiation
reaction force in the classical limit \cite{Landau2}, $\mu=2\omega^{\prime}e^{2}/\left(3mc^{3}\right)$,
$G\left(\chi_{e}\right)=I_{rad}\left(\chi_{e}\right)/I_{rad}\left(\chi_{e}=0\right)$
is the QED factor introduced in order to take into account the decreasing
of the radiation power and the radiation reaction force in the quantum
limit with increasing of $\chi_{e}$ \cite{Bell2008,Bulanov2013,Esirkepov2015}.
For the sake of convenience, hereinafter, the prime symbol are omitted
for the quantities in the HB-frame. 

It is shown for a rotating electric field \cite{Zeldovich1975,Bulanov2011-1}
that there is a stationary trajectory attracting the other trajectories.
The field has to be strong enough so that the electrons and positrons
move in the radiation reaction regime. Regardless of the initial momentum
a positron quickly reaches stationary trajectory which is the rotation
with the field frequency. The phase shift between the field and the
positron velocity is set so that the work done by the electric field
is completely compensated by the radiative losses. 

We extend the Zeldovich model \cite{Zeldovich1975} to the configuration
of the rotating homogeneous electric and magnetic fields, which are
parallel to each other:
\begin{multline}
\mathbf{E}=E_{0}(0,\sin t,\cos t),\mathbf{\: B}=B_{0}(0,\sin t,\cos t).\label{rf}
\end{multline}
The electric and magnetic fields rotate in the plane $y-z$ with the
unit frequency $\omega^{\prime}=1$. Like in the Zeldovich model we
assume that the positron rotates in the plane $y-z$ with the constant
velocity $v_{\perp}$ and frequency $\omega^{\prime}=1$ but it additionally
moves along $x$-axis with the constant velocity $v_{x}$. Balancing
the forces along $x$-axis and in the $y-z$ plane (along the centrifugal
force and along the transversal velocity, respectively) we get
\begin{eqnarray}
\frac{dp_{x}}{dt} & = & v_{\perp}B_{0}\sin\varphi-v_{x}F_{R}=0,\label{z1}\\
\frac{dp_{y}}{dt} & = & E_{0}\sin\varphi+B_{0}v_{x}\cos\varphi=\gamma v_{\perp},\label{z2}\\
\frac{dp_{z}}{dt} & = & E_{0}\cos\varphi-v_{x}B_{0}\sin\varphi-v_{\perp}F_{R}=0,\label{z3}\\
F_{R} & = & \mu G\left(\chi_{e}\right)W^{2}\gamma^{2}\left(1-v_{\perp}^{2}\cos^{2}\varphi\right),\label{z4}\\
\chi_{e} & = & a_{S}^{-1}W\gamma\sqrt{1-v_{\perp}^{2}\cos^{2}\varphi},\\
\frac{d\mathbf{r}}{dt} & = & \frac{\mathbf{p}}{\gamma},\label{z5}
\end{eqnarray}
where it is assumed that $z$-axis is directed along the transverse
component of the positron velocity, $\mathbf{v_{\perp}}$, so that
the centrifugal force is directed along the $y$-axis, $\varphi$
is the angle between $\mathbf{v_{\perp}}$and $\mathbf{E}$, $\gamma^{-2}=1-v_{x}^{2}-v_{\perp}^{2}$
is the reverse squared relativistic Lorentz factor of the positron,
$W^{2}=E_{0}^{2}+B_{0}^{2}$. The first equation represents the balance
between the Lorentz force and the radiation reaction force along $x$-axis,
the second one represents the balance between the centrifugal force
and the Lorentz force. For ultra-relativistic motion $\gamma\gg1$
($v_{x}^{2}\approx1-v_{\perp}^{2}$) Eqs.~(\ref{z1})-(\ref{z4})
can be reduced to the system of equations for $\gamma$, $v_{\perp}$
and $\cos\varphi$:
\begin{eqnarray}
v_{\perp}B_{0} & = & \sqrt{\frac{1-v_{\perp}^{2}}{1-\cos^{2}\varphi}}F_{R}\left(\gamma,v_{\perp},\cos\varphi\right),\label{zsys1}\\
\gamma v_{\perp} & = & E_{0}\sqrt{1-\cos^{2}\varphi}+B_{0}\sqrt{1-v_{\perp}^{2}}\cos\varphi,\label{zsys2}\\
v_{\perp}E_{0}\cos\varphi & = & F_{R}\left(\gamma,v_{\perp},\cos\varphi\right),\label{zsys3}
\end{eqnarray}
where $F_{R}$ is given by Eq.~(\ref{z4}). The third equation can
be derived by summation of Eq.~(\ref{z1}) multiplied by $v_{x}$
and Eq.~(\ref{z3}) multiplied by $v_{\perp}$. It demonstrates that
the radiative losses are completely compensated by the work done by
the electric field, hence $F_{R}\leq E_{0}$. Note that the useful
relations $\gamma v_{\perp}E_{0}=W^{2}\sin\varphi$ and $v_{x}=\left(B_{0}/E_{0}\right)\tan\varphi$
can be derived from Eqs.~(\ref{zsys1})-(\ref{zsys3}). 

In the limit of high field the radiation reaction is strong and $F_{R}\approx E_{0}$,
$\varphi\ll1$, $v_{x}\ll1$ and the solution of Eqs.~(\ref{zsys1})-(\ref{zsys3})
can be written as follows

\begin{eqnarray}
\varphi & \approx & \frac{E_{0}\gamma}{W^{2}}\ll1,\label{zs1}\\
v_{x} & \approx & \frac{B_{0}}{E_{0}}\varphi\ll1,\label{zs2}\\
\chi_{e} & \approx & \frac{\gamma^{2}}{a_{S}},\label{zs3}\\
\gamma=\varepsilon_{e} & \approx & \left[\frac{E_{0}}{\mu G\left(\chi_{e}\right)}\right]^{1/4},\label{zsg}
\end{eqnarray}
where it is assumed that $B_{0}\lesssim E_{0}$. It follows from Eqs.~(\ref{zs1})
and (\ref{zsg}) that the radiation reaction regime corresponds to
the condition $\epsilon_{R}\equiv E_{0}^{3}G\mu\gg1$. In order to
explicitly write expressions for $\varphi$, $v_{x}$ and $\gamma$
we have to solve Eq.~(\ref{zs3}) for $\chi_{e}$
\begin{eqnarray}
\chi_{e}^{2}G\left(\chi_{e}\right) & = & \frac{E_{0}}{\mu a_{S}^{2}}.\label{chie}
\end{eqnarray}
Eqs.~(\ref{zs1}) and (\ref{zsg}) are reduced to the formulas derived
by Zeldovich \cite{Zeldovich1975} in the limit $B_{0}=0$ and $G=1$. 

More accurate value of $\gamma$ (for arbitrary value of $\varphi$
and $\epsilon_{R}$ i.e. not only for the radiation reaction regime)
can be found in the limit $B_{0}\ll E_{0}\epsilon_{R}$ ($v_{x}\ll1$)
from equation 
\begin{eqnarray}
E_{0}^{2}-\gamma^{2}\frac{E_{0}^{2}}{W^{2}} & \approx & \mu^{2}G^{2}\left(\chi_{e}\right)\gamma^{8},\label{zgamma}\\
\chi_{e} & \approx & \frac{\gamma^{2}}{a_{S}}.
\end{eqnarray}
It should be noted that Eq.~(\ref{zgamma}) for $\gamma$ is similar
to one for the electron energy in the rotating electric field \cite{Zeldovich1975}
and in the running circularly polarized wave \cite{Bashinov2014,Brady2014}
for $G\left(\chi_{e}\right)=1$. In the limit for the radiation reaction
regime ($\epsilon_{R}\gg1$) Eq.~(\ref{zgamma}) is reduced to Eq.~(\ref{zsg}).
In the opposite limit, when the radiation reaction can be neglected,
$\gamma\approx W$ and $\gamma\approx E_{0}$ for $B_{0}=0$ in agreement
with the known results  \cite{Zeldovich1975}. 

Combining Eqs.~(\ref{zs3}) and (\ref{zgamma}) the equation for
$\chi_{e}$ can be derived
\begin{eqnarray}
\frac{G^{2}\left(\chi_{e}\right)\chi_{e}^{4}}{W^{2}a_{S}^{-1}-\chi_{e}} & \approx & \frac{E_{0}^{2}}{\mu^{2}W^{2}a_{S}^{3}},\label{zchi}
\end{eqnarray}
In the limit $W^{2}a_{S}^{-1}\gg\chi_{e}$ ($\epsilon_{R}\gg1$) Eq.~(\ref{zchi})
is reduced to Eq.~(\ref{chie}).

The description with the averaged radiation reaction force with QED
factor $G\left(\chi_{e}\right)$ can be used when the number of the
photons emitted during interaction and during the characteristic time
of the field ($1/\omega^{\prime}$ ) is large: $W_{rad}\gg1$, where
$\tau_{rad}\sim W_{rad}^{-1}$ is the characteristic time of the photon
emission. The model is not valid when the electric field is too weak
and Eq.~(\ref{zsys3}) is not fulfilled. In other words, the radiative
losses has to be compensated by the work done by the electric field.
It follows from the obtained result that the stationary trajectory
in the rotating electric and magnetic field is helical so that the
positron drifts along $x$-axis with the constant velocity $v_{x}$
and rotates in $y-z$ plane with phase shift between the field and
the transverse component of the velocity, $\varphi$. The stationary
trajectory $(\mathbf{r}^{z},\mathbf{p}^{z})$ in the the radiation
reaction regime ($\epsilon_{R}\gg1$) can be approximated as follows
\begin{eqnarray}
x^{z} & \approx & v_{x}t,\label{xz}\\
y^{z} & \approx & \cos(t+\varphi),\label{yz}\\
z^{z} & \approx & -\sin(t+\varphi),\label{zz}\\
p_{x}^{z} & = & v_{x}\gamma^{z}\approx\frac{B_{0}}{W^{2}}\gamma^{z},\label{pxz}\\
p_{y}^{z} & \approx & \gamma^{z}\sin(t+\varphi),\label{pyz}\\
p_{z}^{z} & \approx & \gamma^{z}\cos(t+\varphi),\label{pzz}\\
\gamma^{z} & \approx & \left[\frac{E_{0}}{\mu G\left(\chi_{e}\right)}\right]^{1/4},
\end{eqnarray}
where $\varphi$ and $v_{x}$ are given by Eqs.~(\ref{zs1}) and
(\ref{zs2}), respectively.

\begin{figure}[h]
\centering \includegraphics[width=8cm]{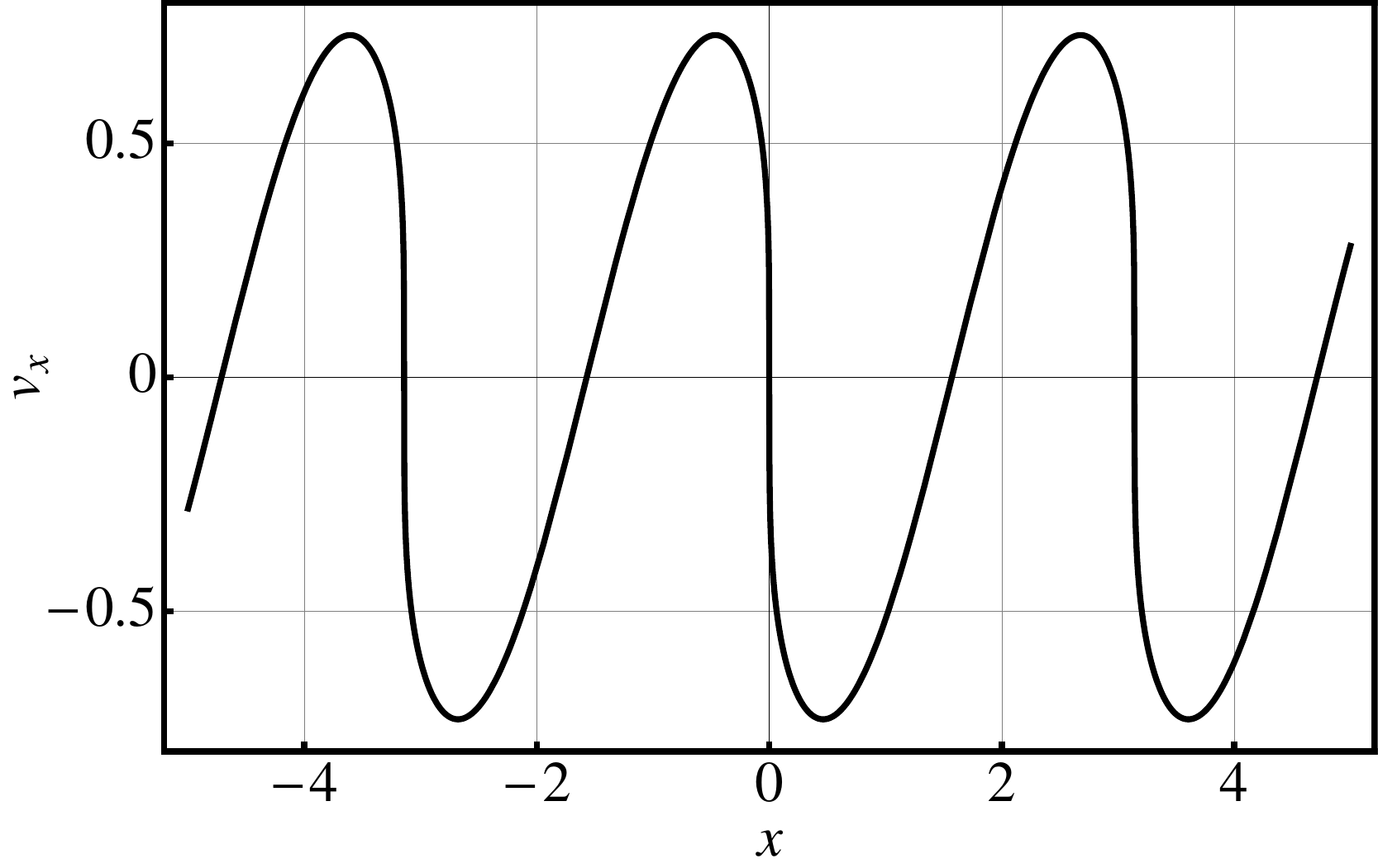} \caption{$v_{x}(x)$ calculated from Eq.~(\ref{vx}) for $u=1$. }

\label{vx-fig} 
\end{figure}

The positron trajectory given by Eqs.~(\ref{zs1})-(\ref{zsg}) is
calculated for homogeneous electric and magnetic fields. However the
solution can be also used to describe the positron motion in the standing
circularly polarized wave far from the electric node (the antinode
of $\mathbf{B}$). This is because of the slow motion of the positron
along $x$-axis ($v_{x}\ll1$) so that the positron has enough time
to switch to the stationary trajectory given by Eqs.~(\ref{xz})-(\ref{pzz})
and determined by the local values of the fields. The trajectory of
the positron created near the magnetic node of the standing wave can
be calculated by taking into account the dependence of $E_{0}$ and
$B_{0}$ on $x$ in Eqs.~(\ref{z1})-(\ref{z5}), where $E_{0}=2a_{0}\sin x$,
$B_{0}=2a_{0}\cos x$ and $W=2a_{0}$. The longitudinal coordinate
can be found from the equation of motion: $dx/dt=v_{x}$ :
\begin{eqnarray}
\intop_{0}^{x^{Z}}\frac{d\xi}{v_{x}\left(E_{0}\left(\xi\right),B_{0}\left(\xi\right)\right)} & = & t,\label{xzsw}
\end{eqnarray}
where $v_{x}$ is the solution of Eqs.~(\ref{zsys1})-(\ref{zsys3}).
$v_{x}$ can be approximated by using of Eq.~(\ref{zs2}) as follows:
\begin{eqnarray}
v_{x} & \approx & u\cos x\left|\sin x\right|^{1/4}\textrm{sign}\left(\sin x\right),\label{vx}
\end{eqnarray}
where $u=\left(8a_{0}^{3}\mu G\left(\chi_{e}\right)\right)^{-1/4}$,
$\chi_{e}$ is the solution of Eq.~(\ref{zchi}) for $E_{0}\approx2a_{0}$,
$\textrm{sign}\left(x\right)=-1$ for $x<0$, $\textrm{sign}\left(x\right)=0$
for $x=0$ and $\textrm{sign}\left(x\right)=1$ for $x>0$. The dependence
of $v_{x}$ on $x$ is shown in Fig.~\ref{vx-fig}. Therefore, the
positron trajectory near the magnetic node of the standing wave takes
a form $\mathbf{r}\approx\mathbf{r}_{0}+\mathbf{r}^{z}(t,E_{0}\left(x\left(t\right)\right),B_{0}\left(x\left(t\right)\right),\varphi\left(x\left(t\right)\right))$,
$\mathbf{p}\approx\mathbf{p}^{z}(t,E_{0}\left(x\left(t\right)\right),B_{0}\left(x\left(t\right)\right),\varphi\left(x\left(t\right)\right))$,
where the constant $\mathbf{r}_{0}$ is determined by the initial
conditions. Note that such constant is absent in the expression for
$\mathbf{p}$ since all positrons locating at the same position on
the stationary trajectory have the same momentum. Evidently, for the
electrons $v_{x}$ is the same as that for the positrons while $\mathbf{v}_{\perp}$
is opposite to that of the positrons.

\begin{figure}[h]
\centering \includegraphics[width=8cm]{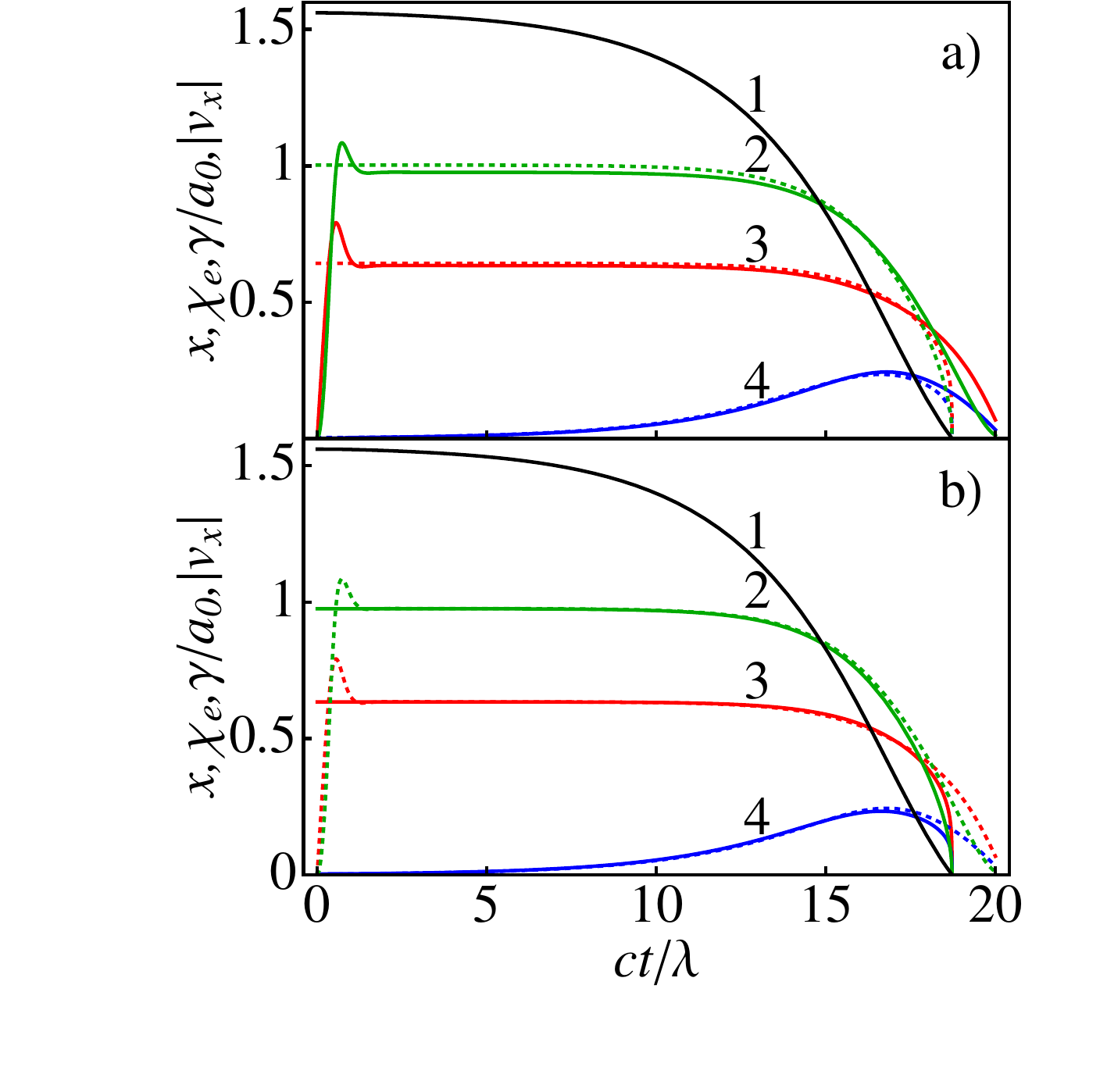} \caption{The approximation $G=1$ (the QED suppression of the radiation reaction
is neglected). $x(t)$ (solid black line 1), $\chi(t)$ (solid green
line 2), $\gamma(t)/a_{0}$ (solid red line 3), $v_{x}(t)$ (solid
blue line 4) calculated numerically by solving Eqs.~(\ref{em1})-(\ref{em3})
for the positron with the initial condition $x(t=0)=0.497\pi$, $p(t=0)=0$
in the standing wave (Eqs.~(\ref{Ehbf}) and (\ref{Bhbf})) with
$a_{0}=1000$. $\chi(x(t))$ (dashed green line 2), $\gamma(x(t))/a_{0}$
(dashed red line 3), $v_{x}(x(t))$ (dashed blue line 4) are calculated
from (a) Eqs.~(\ref{zs1})-(\ref{zsg}) and (b) from Eqs.~(\ref{zs1})-(\ref{zs3}),
(\ref{zgamma}), where $E_{0}=2a_{0}\sin x(t)$, $B_{0}=2a_{0}\cos x(t)$.
$x(t)$ is retrieved from the numerical solution and shown by the
solid black line 1. }

\label{zeld} 
\end{figure}

The equations of motion are solved numerically with the radiation
reaction force for the positron being initially near the magnetic
node ($x(t=0)=0.497\pi$, $p(t=0)=0$) of the standing wave with $a_{0}=1000$.
First we solve Eqs.~(\ref{em1})-(\ref{em3}) neglecting the suppression
of the radiation reaction force ($G=1$). The values of $\gamma(t)$,
$v_{x}(t)$ and $\chi_{e}(t)$ obtained from numerical solution of
equations of motion and ones estimated from Eqs.~(\ref{zs1})-(\ref{zsg})
are shown in Fig.~\ref{zeld}, where in the estimations $E_{0}=2a_{0}\sin x(t)$,
$B_{0}=2a_{0}\cos x(t)$ and $x(t)$ is retrieved from the numerical
solution. It is seen from Fig.~\ref{zeld}(a) that the model prediction
is in a very good agreement with the numerical solution of the equations
of motion. The better agreement is achieved (see Fig.~\ref{zeld}(b))
when Eq.~(\ref{zgamma}) is used instead of Eq.~(\ref{zsg}). It
is interesting to note that even near the electric node ($x\approx0$)
the agreement is still fairly good. 

\begin{figure}[h]
\centering \includegraphics[width=8cm]{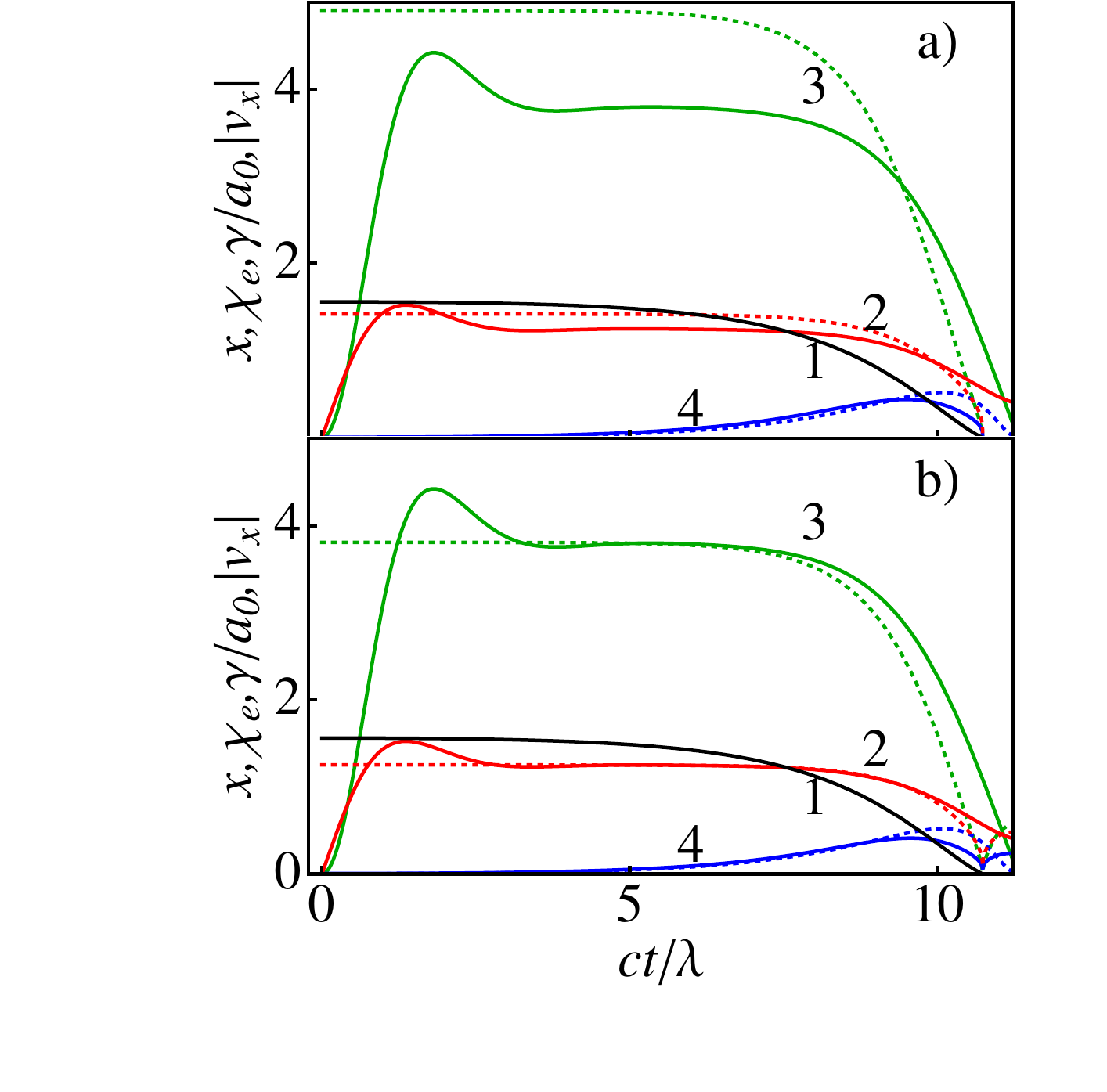} \caption{The case when the QED suppression of the radiation reaction is taken
into account. $x(t)$ (solid black line 1), $\gamma(t)/a_{0}$ (solid
red line 2), $\chi(t)$ (solid green line 3), $v_{x}(t)$ (solid blue
line 4) calculated numerically by solving Eqs.~(\ref{em1})-(\ref{em3})
for the positron with the initial condition $x(t=0)=0.497\pi$, $p(t=0)=0$
in the standing wave (Eqs.~(\ref{Ehbf}) and (\ref{Bhbf})) with
$a_{0}=1000$. $\chi(x(t))$ (dashed green line 3), $\gamma(x(t))/a_{0}$
(dashed red line 2), $v_{x}(x(t))$ (dashed blue line 4) are calculated
from (a) Eqs.~(\ref{zs1})-(\ref{zsg}), (\ref{chie}) and (b) from
Eqs.~(\ref{zs1})-(\ref{zs3}), (\ref{zchi}), where $E_{0}=2a_{0}\sin x(t)$,
$B_{0}=2a_{0}\cos x(t)$. $x(t)$ is retrieved from the numerical
solution and shown by the solid black line 1. }

\label{zeld-bula} 
\end{figure}

\begin{figure}[h]
\centering \includegraphics[width=8cm]{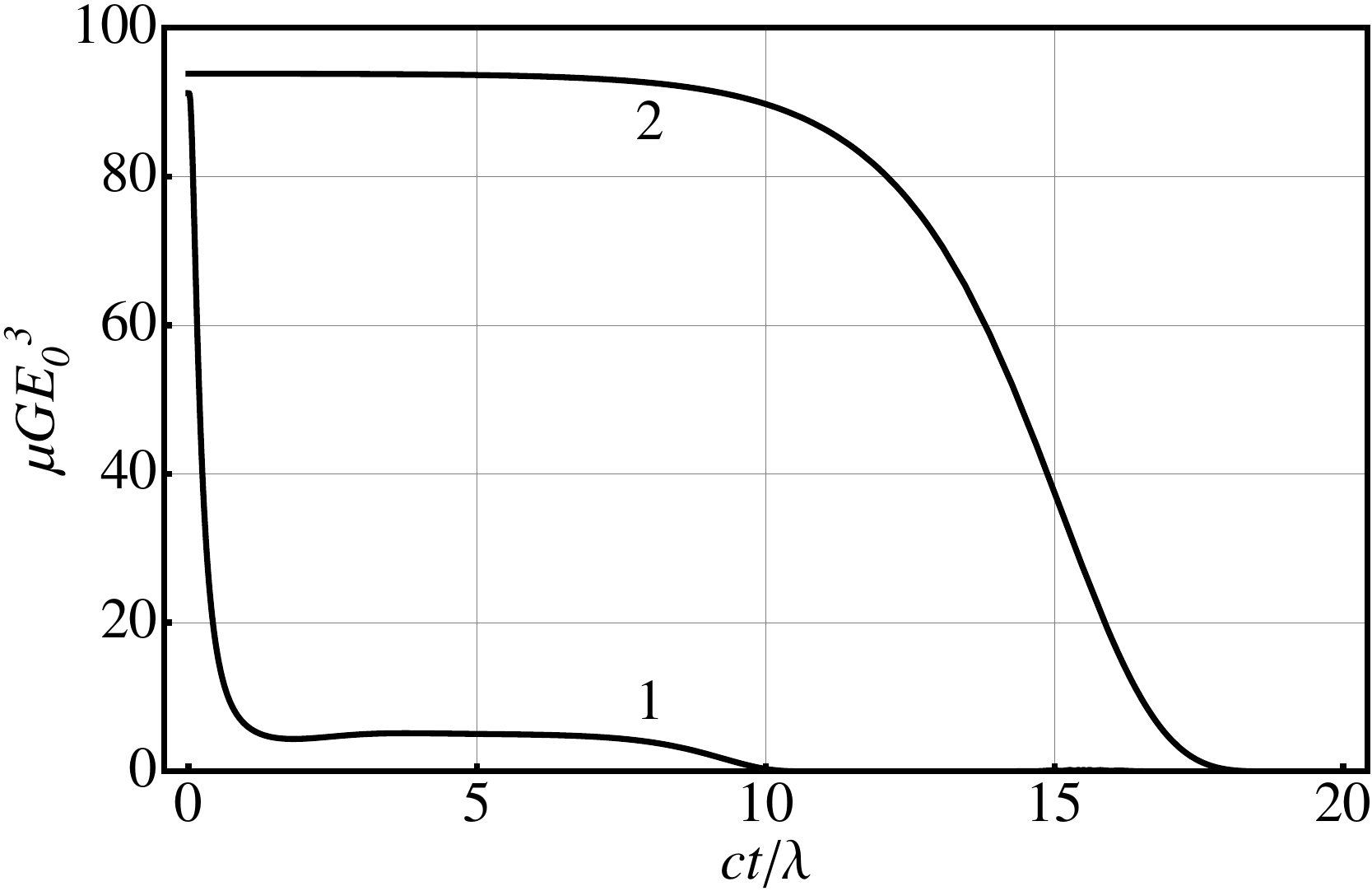} \caption{The radiation reaction parameter $\epsilon_{R}(t)=\mu G\left(\chi_{e}\left(t\right)\right)E_{0}^{3}\left(t\right)$
for the positron with the initial condition $x(t=0)=0.497\pi$, $p(t=0)=0$
in the standing wave (Eqs.~(\ref{Ehbf}) and (\ref{Bhbf})) with
$a_{0}=1000$ in the case when the QED suppression of the radiation
reaction is taken into account (line 1) and in the approximation $G=1$
when the suppression is neglected (line 2). }

\label{muchiE3} 
\end{figure}

We also solve the equations of motion numerically for the positron
with the same initial condition taking into account QED suppression
of the radiation reaction force, where the approximation $G(\chi_{e})\approx\left(1+18\chi_{e}+69\chi_{e}^{2}+73\chi_{e}^{3}+5.804\chi_{e}^{4}\right)^{-1/3}$
proposed in Ref.~\cite{Esirkepov2015} is used. The values of $\gamma(t)$,
$v_{x}(t)$ and $\chi_{e}(t)$ obtained from numerical solution of
equations of motion and ones estimated from Eqs.~(\ref{zs1})-(\ref{zs3}),
(\ref{chie}) are shown in Fig.~\ref{zeld-bula}(a) and ones estimated
from Eqs.~(\ref{zs1})-(\ref{zs3}), (\ref{zchi}) are shown in Fig.~\ref{zeld-bula}(b).
In the estimations $E_{0}=2a_{0}\sin x(t)$, $B_{0}=2a_{0}\cos x(t)$,
where $x(t)$ is retrieved from the numerical solution of the equations
of motion. It is seen from Fig.~\ref{zeld-bula}(a) that the agreement
between the quantities calculated numerically and the estimated ones
is not so good as in the case $G=1$. The discrepancy is caused by
strong radiation reaction suppression ($G\ll1$). It is seen from
Fig.~\ref{muchiE3} that the radiation reaction parameter $\epsilon_{R}(t)=\mu G\left(\chi_{e}\left(t\right)\right)E_{0}^{3}\left(t\right)$
determining the transition to the radiation reaction regime decreases
in about $20$ times when the suppression is taken into account. In
this case the parameter is close to $5$ and is not sufficient to
ensure the required accuracy of the approximation corresponding to
the radiation reaction regime and described by Eqs.~(\ref{zs1})-(\ref{zs3}),
(\ref{chie}). Significant improvement of the accuracy can be achieved
when more general Eq.~(\ref{zchi}) is used instead of Eq.~(\ref{chie})
(see Fig.~\ref{zeld-bula}(b)). 

It follows from Fig.~\ref{zeld-bula} that $\gamma\sim1200$ and
$\chi_{e}\sim4$ in the case when the radiation reaction suppression
is included. The positron energy and the parameter $\chi_{e}$ are
higher in several times than ones in the case $G=1$ (see Fig.~\ref{zeld}).
The probability rate for photon emission given by Eq.~(\ref{Wr1})
is $W_{rad}\sim8$ for $\gamma\sim1200$ and $\chi_{e}\sim4$. Therefore
the positron passing from the magnetic node to the electric one during
$t_{int}\sim11$ emits about $N_{ph}\sim t_{int}/\tau_{rad}\sim t_{int}W_{rad}\sim90\gg1$
photons and the approximation with the averaged radiation reaction
force by means of factor $G$ can be applied. 

It follows from Eqs.~(\ref{vx}) that the longitudinal velocity of
the positrons and electrons created with small momentum in the standing
circularly polarized wave is directed from the magnetic nodes ($x=\pm\pi\left(n+1/2\right)$,
$n=0,1,2,...$, where $\mathbf{B}=0$ and the electric field amplitude
peaks) to the electric ones ($x=\pm\pi n$, $n=0,1,2,...$, where
$\mathbf{E}=0$ and the magnetic field amplitude peaks). Hence, the
magnetic nodes are unstable for the positrons and electrons while
the electric nodes are stable for them (see Fig.~\ref{vx-fig}).
In the magnetic nodes the positrons and electrons perform circular
motion in the rotating electric field. In the electric nodes the positrons
and electrons move in the rotating magnetic field. This motion is
complex and can be qualitatively presented as the superposition of
the fast cyclotron rotation (rotation axis is perpendicular to the
$x\mbox{-axis)}$ and slow drift. The frequency of the cyclotron rotation
in the magnetic field is much higher than the field frequency $\omega_{B}\approx2a_{0}\omega^{\prime}/\gamma\gg\omega^{\prime}$
since the positrons and electrons move in the radiation reaction regime
for $a_{0}>300$ so that $\gamma/a_{0}\sim\epsilon_{R}^{-1/4}\ll1$
(see Eq.~(\ref{zsg}) and Refs.~\cite{Bashinov2014,Brady2014,Zeldovich1975}).

When the number of the electron-positron pairs becomes large they
produce more photons than ones arrived from the electron layer. As
a result the self-sustained QED cascade characterized by exponential
growth of the pair number in time can develop. It is demonstrated
\cite{Bashmakov2014} that the cascade growth rate is maximal in the
magnetic nodes of the circularly polarized standing wave. However
it is discussed above that the the pair positions is unstable in the
magnetic nodes and is stable in the electric ones. The pair density
profile is determined by the trade off between the pair production
effect and the pair drift. Therefore the density of the electron-positron
plasma may peak at the electric and magnetic nodes as the pairs production
is the most efficient at the magnetic nodes while the pairs after
creation are attracted to the electric nodes.

\section{Numerical simulations}

\begin{figure}[h]
\centering \includegraphics[width=8cm]{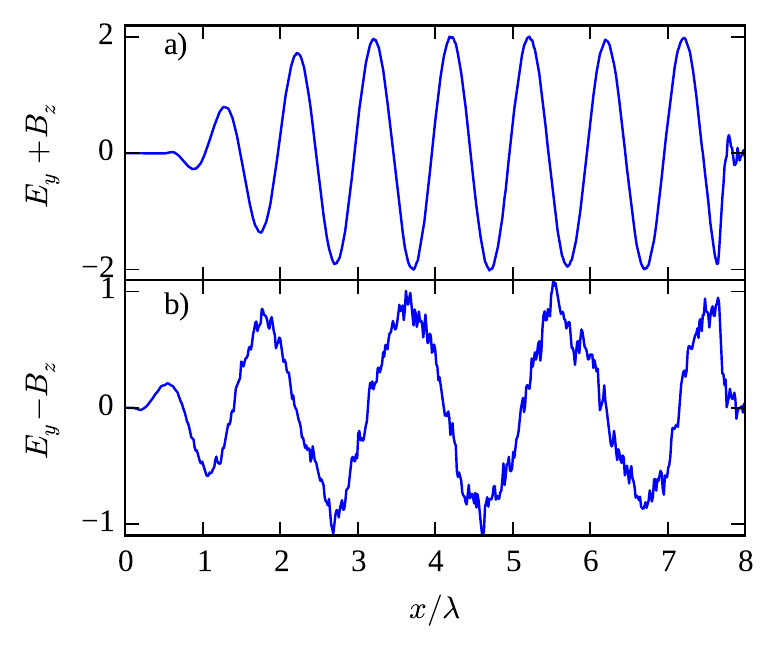} \caption{(a) $E_{y}+B_{z}$ and (b) $E_{y}-B_{z}$ as a function of $x$ in
front of the foil for $a_{0}=1000$ at $t=6\lambda/c$ .}

\label{fields1000} 
\end{figure}

\begin{figure}[h]
\centering \includegraphics[width=8cm]{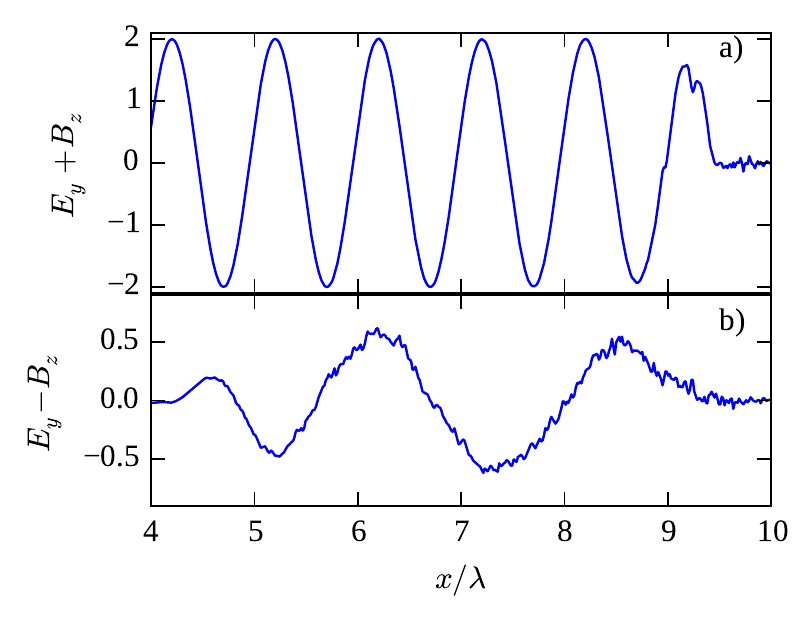} \caption{(a) $E_{y}+B_{z}$ and (b) $E_{y}-B_{z}$ as a function of $x$ in
front of the foil for $a_{0}=1840$ at $t=4\lambda/c$ .}

\label{fields1840} 
\end{figure}

The field of the incident wave can be retrieved from $E_{y}+B_{z}$
and the field of the reflected wave can be retrieved from $E_{y}-B_{z}$
calculated in the numerical simulations. To compare the analytical
results with the numerical ones it is convenient to use other dimensionless
units, normalizing the time to $1/\omega_{L}$ , the length to $c/\omega_{L}$
, and the field amplitude to $mc\omega_{L}/e$. It follows from Eqs.~(\ref{CPinLF}),
(\ref{CPinLF2}), (\ref{ErnLF}), (\ref{BRinLF}) that
\begin{align}
E_{i,y}+E_{r,y}+B_{i,z}+B_{r,z} & =2a_{0}\cos\left(x-t\right),\label{FLF1}\\
E_{i,y}+E_{r,y}-B_{i,z}-B_{r,z} & =2a_{0}\omega_{r}\nonumber \\
\times & \cos\left[\omega_{r}\left(x+t\right)\right].\label{FLF2}
\end{align}
The HB front velocity and the frequency of the reflected wave can
be estimated by using Eqs.~(\ref{vhb}), (\ref{mu}) and (\ref{wr}).
Then for the simulation parameters we get: $\mu=1.84$, $v_{HB}\approx0.35$,
and $\omega_{r}\approx0.48$ for $a_{0}=1000$, while $\mu=1$, $v_{HB}=1/2$
and $\omega_{r}\approx0.33$ for $a_{0}=1840$ that is close to the
simulation results, namely form the periods of the wave $E_{y}-B_{z}$
(see Figs.~\ref{fields1000} and \ref{fields1840}) we obtain $\omega_{r}\approx0.5$
for $a_{0}=1000$ and $\omega_{r}\approx0.4$ for $a_{0}=1840$. According
to the model assumptions the refection in the HB-frame is perfect
so that the reflection coefficient in the laboratory frame is equal
to $r=\max\left[(E_{y}-B_{z})/(E_{y}+B_{z})\right]=\omega_{r}$. This
is also in a good agreement with the results of the numerical simulations
(see Figs.~\ref{fields1000} and \ref{fields1840}). Therefore the
approximation of the structure of the electromagnetic field in the
vacuum region (in front of the foil) as a standing wave can be used
for estimations. 

\begin{figure}[h]
\centering \includegraphics[width=8cm]{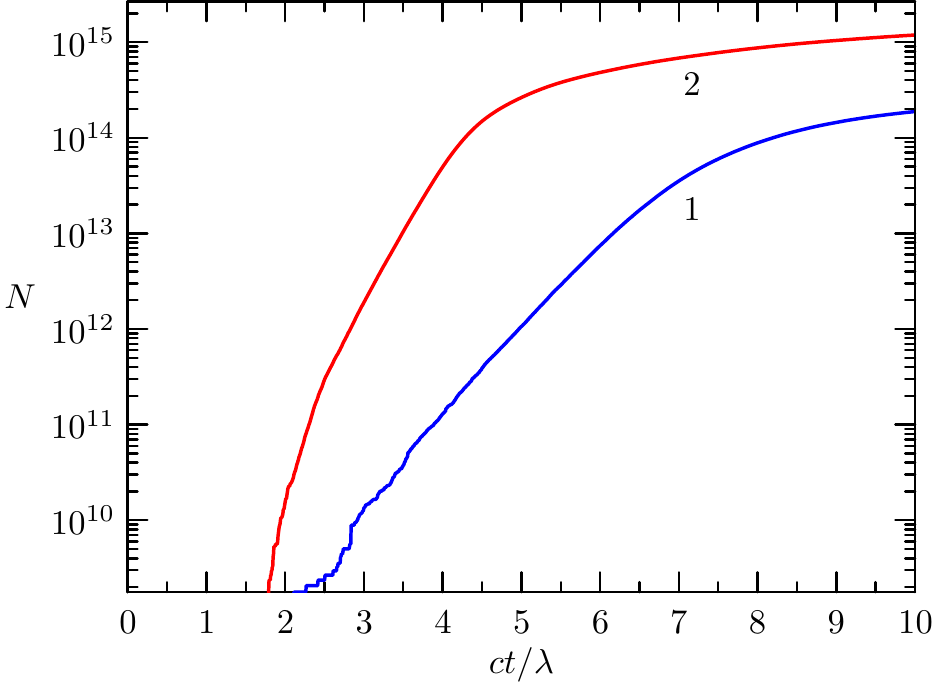} \caption{Positron number as a function of time for $a_{0}=1000$ (line 1) and
$a_{0}=1840$ (line 2). }

\label{N} 
\end{figure}

The positron number as a function of time is shown in Fig.~\ref{N}.
It follows from Fig.~\ref{N} that the exponential growth representing
QED cascading starts almost from the beginning. The cascade develops
in the circularly polarized standing wave generated in the HB-frame
in front of the foil. The cascade growth rate can me estimated from
the figure: $\Gamma\approx0.6$ for $a_{0}=1000$ and $\Gamma\approx1.3$
for $a_{0}=1840$, where the cascade growth rate is normalized to
the frequency of the standing wave in the HB-frame, $\omega^{\prime}=\omega_{L}\gamma_{HB}\left(1-v_{HB}\right)$.
The obtained values of $\Gamma$ are slightly less than that calculated
in Ref.~\cite{Grismayer2016} by numerical simulation for the rotating
electric field and for the circularly polarized standing wave ($\Gamma\approx0.8$
for $a_{0}=1000$ and $\Gamma\approx1.8$ for $a_{0}=1840$, see Fig.~2a
in Ref.~\cite{Grismayer2016}). The reason is that the standing wave
is not perfect in our case because the laser radiation reflection
from the foil is not also perfect. 

\begin{figure}[h]
\centering \includegraphics{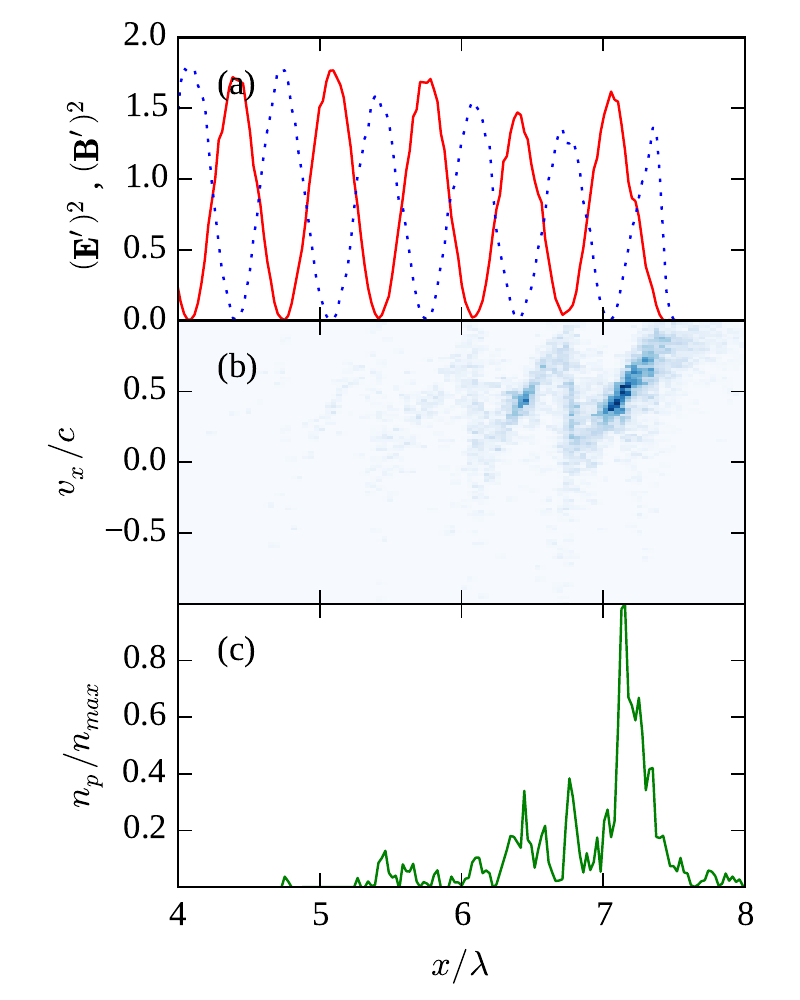} \caption{(a) The squared electric and magnetic fields in the HB-frame, $\left(\mathbf{E}^{\prime}\right)^{2}$(solid
red line) and $\left(\mathbf{B}^{\prime}\right)^{2}$ (dashed blue
line), as a function of $x$, (b) the positron distribution in the
plane $x-v_{x}$ and (c) the positron density along the $x$-axis
as a function of $x$ for $a_{0}=1000$ and $t=6.4\lambda/c$. }

\label{a1000} 
\end{figure}

\begin{figure}[h]
\centering \includegraphics{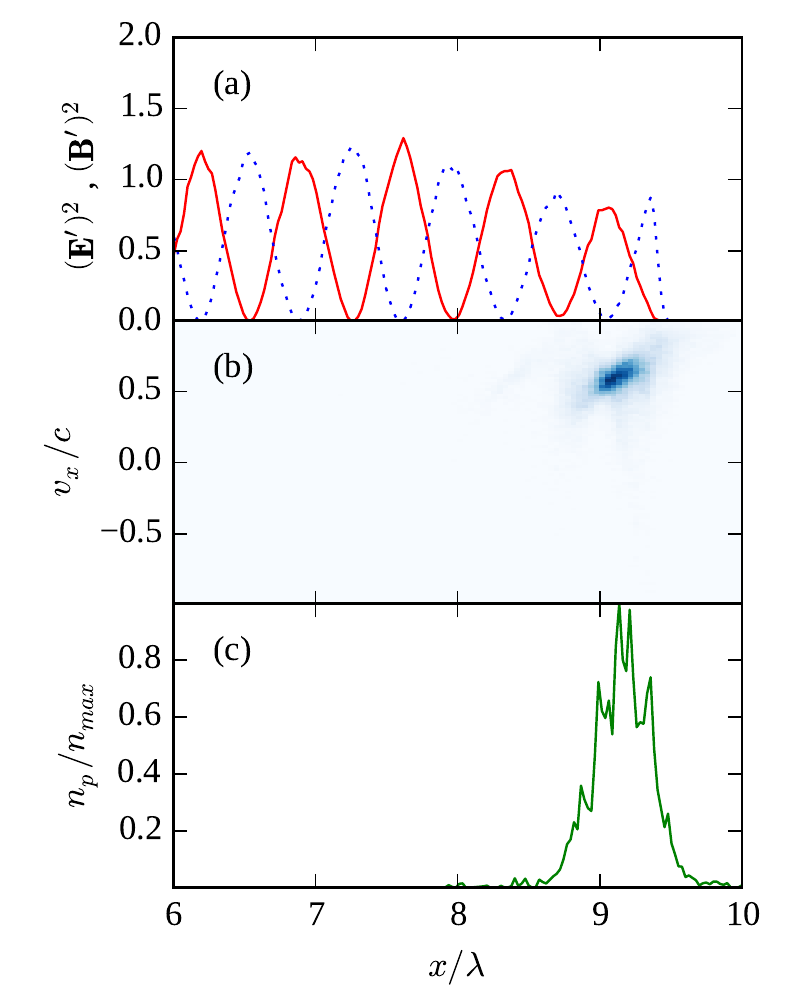} \caption{(a) The squared electric and magnetic fields in the HB-frame, $\left(\mathbf{E}^{\prime}\right)^{2}$(solid
red line) and $\left(\mathbf{B}^{\prime}\right)^{2}$ (dashed blue
line), as a function of $x$, (b) the positron distribution in the
plane $x-v_{x}$ and (c) the positron density along the $x$-axis
as a function of $x$ for $a_{0}=1840$ and $t=4.0\lambda/c$. }

\label{a1840} 
\end{figure}

According to Eq.~(\ref{Ehbf}) the square of the electric field in
the vacuum region in the HB-frame as a function of the space-time
position in the laboratory frame takes a form: 
\begin{align}
\left(\mathbf{E}^{\prime}\right)^{2} & =\left(2a_{0}\omega^{\prime}\right)^{2}\text{\ensuremath{\sin}}^{2}\left(\omega^{\prime}x^{\prime}\right)\nonumber \\
 & =\left(2a_{0}\right)^{2}\frac{1-v_{HB}}{1+v_{HB}}\text{\ensuremath{\sin}}^{2}\left(\frac{x-v_{HB}t}{1+v_{HB}}\right).
\end{align}
To calculate the fields in the HB-frame we can apply Lorentz transformation
to the field distribution retrieved from the numerical simulations
with $v_{HB}\approx0.35$ for $a_{0}=1000$ and $v_{HB}=0.5$ for
$a_{0}=1840$. Thus the positions of nodes and antinodes in the laboratory
reference frame can be easily found from the distribution of $\left(\mathbf{E}^{\prime}\right)^{2}$.
The squared electric and magnetic fields in the HB-frame, $\left(\mathbf{E}^{\prime}\right)^{2}$
and $\left(\mathbf{B}^{\prime}\right)^{2}$, as a function of $x$,
the positron distribution in the plane $x-v_{x}$ and the positron
density at the axis $y=z=0$ as a function of $x$ are shown in Fig.~\ref{a1000}
for $a_{0}=1000$ and $t=6.4\lambda/c$ and in Fig.~\ref{a1840}
for $a_{0}=1840$ and $t=4.0\lambda/c$. It is seen from Fig.~\ref{a1840}
that for strong laser field with $a_{0}=1840$ the most of the positrons
are created in front of the foil near the first magnetic node of the
standing wave because the most of the photons, which are emitted from
the foil and initiating the cascade, decay already in the first period
of the standing wave. In this case the pair production effects dominate
over the pair drift so that the number of the pairs produced at the
magnetic nodes is higher than that drift to the electric nodes \cite{Bashmakov2014}.
In other words, the time of the particle doubling is less than the
time which takes for the the particles to pass from a magnetic node
to the neighboring electric nodes. For $a_{0}=1000$ the probability
of the pair production is lower than for $a_{0}=1840$ and the positrons
are located in the several wavelengths in front of the foil near the
electric and magnetic nodes (see Fig.~\ref{a1000} (c)) that is in
the qualitative agreement with the predictions formulated in the previous
Section. Small shift of the maximums of the density profile from the
exact position of the nodes can be caused by fact that the reflection
is not perfect so that the wave in the HB-frame is not exactly standing. 

It follows from Figs.~\ref{a1000} and \ref{a1840} that the longitudinal
velocity of the positrons is close to the HB front velocity at the
magnetic nodes and the velocity distributed within wide range near
the electric node. For $a_{0}=1000$ the positron distribution is
sawtooth-like in $x-v_{x}$ plane (see Fig.~\ref{a1000}(b)). Therefore,
in the HB front frame, the longitudinal positron velocity increases
towards the HB front from one electric node to another electric node
reaching $v_{x}=0$ at the magnetic nodes. This is in qualitative
agreement with Eq.~(\ref{vx}) describing sawtooth-like distribution
(see Fig.~\ref{vx-fig}). The longitudinal dynamics of the secondary
electrons is the same as that of the positrons. 

\section{Discussions and conclusions}

It is demonstrated that numerous electron-positron pairs are produced
in the hole-boring regime of interaction between a foil and laser
pulse with intensities higher than $10^{24}$W cm$^{-2}$. The pair
production scenario can be roughly divided into three stages: (i)
cascade initiation by the photons emitted from the foil electrons;
(ii) self-sustained QED cascading in the standing wave; (iii) the
back reaction of the produced pair plasma on the laser-foil interaction.
In the first two stages the pairs are mainly located in the vacuum
region in front of the foil where the incident and reflected laser
waves interfere. When the number of the produced electron-positron
pairs is not very large the field structure in the vacuum region is
close to the standing circularly polarized wave in the hole-boring
front reference frame. The electron-positron plasma is mainly produced
as a result of QED cascading in the standing wave. 

The analytical model for the dynamics of the electrons and positrons
in the rotating electric field with radiation reaction is extended
to the rotating electric and magnetic fields which are parallel to
each other. The model proposed by Zeldovich \cite{Zeldovich1975}
predicts the stationary trajectory attracting the electron trajectories
in the rotating electric field when the radiation reaction is strong.
On such trajectory the work done by the electric field is balanced
by the radiative losses. The particle performs circular motion and
the energy balance is controlled by the phase shift between the electric
field and the particle velocity. In the case of the rotating electric
and magnetic fields, which are parallel to each other, the stationary
trajectory also exists and is helical-like with infinite motion along
the axis perpendicular to the plane of the field rotation. The dynamics
in the circularly polarized standing wave is more complex \cite{Lehmann2012,Esirkepov2015}.
Moreover, the particle motion can be stochastic and the attractor
accumulating the trajectories in the phase space is located only at
the electric node. The attractors are recently studied in the field
of standing waves of various configurations \cite{Lehmann2012,Gonoskov2014,Esirkepov2015,Kirk2016}.
Our model allows one to calculate the particle trajectory near the
magnetic node of the standing wave. It is shown that the trajectory
of the electron and positrons near the magnetic node is close to the
stationary trajectory in the local electric and magnetic fields. The
model includes the QED effect of the radiation reaction suppression
because of the reduction of the total power radiated by the particle
in the quantum regime \cite{Bell2008,Bulanov2013,Esirkepov2015}.

The calculated trajectories are used to analyze the positron density
distribution in the standing wave. It follows from the model that
the positron density peaks at the nodes and antinodes of the standing
wave because the electron-positron pairs are mainly produced at the
magnetic nodes as the cascade growth rate peaks there and the produced
pairs drift to the electric nodes as the magnetic nodes are unstable
for them. The positron distribution in $x$-$v_{x}$ plane is sawtooth-like
and the longitudinal velocity of the positrons is equal to HB-front
velocity at the magnetic nodes. Near the electric nodes the motion
of the electrons and positrons is close to the superposition of the
drift and the rotation so that the longitudinal velocity varies within
the wide range. This is in agreement with the results of the numerical
simulations. 

In the case of high laser intensity ($a_{0}=1840$) the density peaks
at the magnetic node closest to the HB front. The reason is that the
high-energy photons emitted by the foil electrons decay rapidly and
cannot initiate cascade far from the HB-front. The number of the pairs
in the in the electric nodes is much smaller than that in the magnetic
ones because near the magnetic node the pair production rate dominates
over the pair loss rate due to the drift. 

The first stage representing the cascade initiation is not pronounced
in Fig.~\ref{N}. One of the reason is that the number of the high-energy
photons emitted by the electron layer is not very large because the
laser field is strongly suppressed in the layer and the layer electrons
are not accelerated so efficiently as the positrons and the secondary
electrons in the vacuum region in front of the foil. Therefore the
number of the high-energy photons emitted by the pairs and participating
in cascading will exceed the number of the photons emitted by the
foil electrons in very short period of time so that the duration of
the first stage may be small. 

When the pair number becomes great the produced electron-positron
plasma can absorb the laser radiation and affect the dynamics of the
laser-foil interaction. The manifestation of such nonlinear stage
(the third stage) can been seen in Figs.~\ref{a1000}(a) and \ref{a1840}(a)
where the standing wave is slightly attenuated towards the HB front.
The transition between the second and the third stage can be seen
in Fig.~\ref{N} as a saturation of the pair number growth. The transition
occurs at $t\approx7.5\lambda/c$ for $a_{0}=1000$ and for $t\approx4.5\lambda/c$
for $a_{0}=1840$. The analytical model of the third stage has been
proposed in Ref.~\cite{Kirk2013}. It is based on one-dimensional
solutions of the two-fluid (electron-positron) and Maxwell equations,
including a classical radiation reaction term. The model predicts
the vacuum gap with the standing wave structure between the pair ``cushion''
and the targets. However the model verification by self-consistent
numerical simulations is still absent and the detailed analysis of
the third stage with back reaction is needed. 

\begin{acknowledgments} 

This work was supported by the Russian Science Foundation Grant No.
16-12-10383. 

\end{acknowledgments}

\appendix

\end{document}